\documentstyle[pre,aps,epsfig]{revtex}
\begin{document}
\draft
\title{Discrete kink dynamics in hydrogen-bonded chains I:
The one-component model}
\author{V.~M.~Karpan$^{1,2}$, Y.~Zolotaryuk$^{1,2}$,
P.~L.~Christiansen$^1$, and A.~V.~Zolotaryuk$^{1,2}$
}
\address{$^1$Section of Mathematical Physics, IMM, Technical
University of Denmark, DK-2800 Lyngby, Denmark \\
$^2$Bogolyubov Institute for Theoretical Physics,
03143 Kyiv, Ukraine}

\date{\today}

\wideabs{
\maketitle
\begin{abstract}
We study topological solitary waves (kinks and antikinks)
in a nonlinear one-dimensional Klein-Gordon chain
with the on-site potential of a double-Morse type.
This chain is used to describe the collective proton dynamics
in quasi-one-dimensional networks of hydrogen bonds,
where the on-site potential plays role of the proton potential
in the hydrogen bond.
The system supports a rich variety of stationary kink solutions
with different symmetry properties.
We  study the stability and bifurcation structure of
all these stationary kink states. An exactly solvable
model with a piecewise ``parabola-constant'' approximation
of the double-Morse potential is suggested and studied
analytically.
The dependence of the Peierls-Nabarro potential on the system
parameters is studied. Discrete travelling-wave solutions of
 a narrow permanent profile are shown to exist,
depending on the anharmonicity of the Morse potential
and the cooperativity of the hydrogen bond (the coupling constant of 
the interaction
between nearest-neighbor protons).
\end{abstract}
}

\section{Introduction}

Hydrogen bonds (H-bonds) play a crucial role
in the structure and
the dynamics in a whole variety of systems ranging from
ferroelectrics to biomolecules. They are of central importance
in biology, when reactions are considered at molecular level.
In bioenergetics, they appear even more crucial because they enable
transfers of protons from one molecule to another one
in networks or  chains formed via hydrogen bonding \cite{ntn}.

More specifically, H-bonds \cite{vl}  are
interactions linking two molecules or ions, for example, O, N, F,
and Cl atoms,  or in general any pair of
hydroxyl groups, which  may be denoted by X,
 via a hydrogen ion (proton) H$^+$, forming  a 
hydrogen-bonded (HB) bridge
 X--H$\cdots$X as shown schematically in Fig.~\ref{fig1}.
The ion to which the proton in this bridge (H-bond)
 is more tightly linked is called
the hydrogen donor, whereas the other ion is the hydrogen 
acceptor.  More precisely, the proton
is
coupled to each X$^-$ ion through a pair ion-proton
interaction potential of
the standard type (Morse, Lennard-Jones, etc.)
with an equilibrium distance $r_0$, which
 necessarily has a  finite dissociation energy as the
X$\cdots$H distance tends to infinity. 
%
%

As usual, the total 
potential for the HB proton is of a double-well shape, but this
can occur only if the motion of the heavy ions along the H-bond
is appropriately constrained, so that either (i) a sufficiently
strong interaction (repulsion) between the X$^-$ ions 
that does not allow the ions to get closer
each other  a distance less or equal to
$2r_0$ or  (ii) a periodic substrate potential with period
exceeding  $2r_0$ is  additionally involved. In this way, 
a  one-dimensional network of hydrogen bonds can be formed
as a {\it diatomic} chain of alternating heavy (ion) and 
light (proton) particles coupled nonlinearly (e.g., via a
 Morse-type potential, like in Ref.~\cite{c-l}), whereas the 
second-neighbor (ion-ion and proton-proton) interactions
are involved in the harmonic approximation. Under certain
conditions on the ion-ion coupling discussed in Ref.~\cite{zps}, 
 the proton
in each H-bond of the HB chain can be found in
two equilibrium positions
separated by a potential barrier, so that
 the two degenerate ground states  of the chain
$\cdots$ X--H $\cdots$ X--H $\cdots$ X--H $\cdots$
and $\cdots$ H--X $\cdots$ H--X $\cdots$ H--X $\cdots$
are assumed to exist.
Another important property, more specific for biological
systems, is that
 the height of the potential  barrier crucially depends on
the distance between adjacent X$^-$ ions.
Using these properties as main features of  HB chains,   
a number of one-dimensional {\it two-sublattice}  models, whose 
dynamic behavior is governed by the {\it soliton} theory
\cite{zps}, has been suggested and studied extensively.
These soliton-like theories are based on the well-known
{\it cooperativity } of the hydrogen bonding, simply defined
through the coupling of protons in the nearest-neighbor
hydrogen bridges of the chain.

Since the HB chain is a  diatomic lattice, the mechanism of 
hydrogen bonding involves
two  types of  particle displacements. 
Let $Q_n$ and $q_n$ be the displacements of the heavy ion 
and the proton in the $n$th unit cell of the lattice from their
equilibrium positions, at which one of the two ground states
of the chain is realized, respectively. These displacements 
are labeled according to the sequence 
$\{ \ldots , Q_{n-1}, q_{n-1}, Q_n , q_n , Q_{n+1}, q_{n+1} ,
\ldots \}$. Then the general and the most simple
model for the proton transfers in such a diatomic chain
can be given through the  two-sublattice
 Hamiltonian, consisting of two parts \cite{zps}:
\begin{equation}
H= H_0 + H_{ion} .
\label{1}
\end{equation}
The first part
\begin{equation}
H_0=\sum_n\left[\frac{m_p}{2}\dot{q}_n^2 +
\frac{K_p}{2}(q_{n+1}-q_n)^2 + \varepsilon_0 V(u_n, \rho_n)
\right] ,
\label{2}
\end{equation}
with
\begin{equation}
u_n=q_n -\frac{1}{2}(Q_n +Q_{n+1}) ~~\mbox{and}~~
\rho_n =Q_{n+1}-Q_n ,
\label{3}
\end{equation}
describes the proton kinetic energy, the nearest-neighbor
proton-proton interaction, and the intrabond proton energy
that depends on the 
displacements of the protons from the midpoints  in the 
H-bonds $u_n$'s and the relative distances  between the
nearest-neighbor ions $\rho_n$'s, 
whereas the second (pure heavy-ion)
part
\begin{equation}
H_{ion}=\sum_n\left[\frac{1}{2}M\dot{Q}_n^2
 + \frac{1}{2}K_{ion} \rho_n^2
+\frac{1}{2}K_{sub} Q_n^2 \right],
\label{4}
\end{equation}
describes  the kinetic energy of the X$^-$ ions,
the coupling energy between the nearest-neighbor ions,
and the interaction energy  of the X$^-$ ions with
a possible substrate (e.g., formed by the walls of a pore
crossing a membrane).
Here the overdot denotes differentiation on time $t$.
 The proton and ion masses are denoted by $m_p$ and $M$,
respectively. Similarly,
$K_p$, $K_{ion}$, and $K_{sub}$ stand for the stiffness constants
of the interaction between the nearest-neighbor protons, 
the nearest-neighbor ions, and the
chain
ions with the substrate, respectively. It is important that
the intrabond proton energy is given in terms of a general
double-well potential  $V(u,\rho)$  as a function
of two variables: $u$, the proton displacement from
the middle of the hydrogen bond, and $\rho$, the relative 
ion displacement. If additionally this (dimensionless) 
function is normalized according to the relations
$V(0,0)=1$ and $V(\pm a, 0)=0$, where $\pm a$ are
the positions of minima in this function, then
$\varepsilon_0$ is the barrier height of the proton
potential in the H-bond.  When the heavy ions are displaced 
from their equilibria, this potential is deformed, with its barrier 
top moved together with the ions. 


There have been numerous studies \cite{zps} of  soliton
solutions to the equations of motion governed by 
the Hamiltonian  (1)-(4)  including also 
one-component models, where the heavy ions X$^-$
 are
assumed to be fixed \cite{kkn,pn,kr,kez,p-v,m-y,k-z}.
 All these studies refer to the
continuum limit, which presumes the existence
of a
sufficiently effective cooperativity of hydrogen bonding or,
in other words, the
inter-bond proton-proton coupling $K_p$ 
is required to be strong enough.
 However, according to the
{\it ab initio} calculations of the proton-proton 
interaction in realistic HB chains 
by Godzik \cite{go}, 
 $K_p \simeq 41$  Kcal/mol~\AA$^2$. This magnitude appears not
to be sufficient for a {\it free} propagation of the ionic defects
along the HB chain with realistic values of the 
potential barrier height $\varepsilon_0$. The reduction of this barrier
on the basis of the two-component modelling was also shown to be
 not enough to provide a free soliton regime and the ionic defects
in HB chains with realistic parameter values \cite{ho}
were shown to be very narrow objects \cite{sz}.

On the other hand, a rapidly increasing number of publications
 over last years (since the pioneering work of
Peyrard and Kruskal \cite{pk}) have demonstrated significant
differences in the behavior of soliton solutions treated
in the continuum limit and their spatially
discrete relatives \cite{bk}.
The discrete versions of the partial differential equations
brings about a number of critically important modifications
to the dynamics. The moving kinks of the continuum theories
become propagating structures that decelerate by emitting
radiation as they traverse the lattice sites. This ultimately
brakes the structures and brings them to rest, or ``pins''
 them.

The above results has been obtained for the conventional models
such as the discrete sine-Gordon and $\phi^4$ chains. 
For more general
class of models, some interesting and intriguing 
results has been obtained.
Thus, it has been shown that the
{\it shape} of the on-site (in our case, the intrabond proton 
energy) potential is a factor of
particular importance for modelling soliton motion in 
physical systems.
To study this effect,
Peyrard and Remoissenet \cite{pr82prb} have introduced a modified
sine-Gordon
system, where the shape of the on-site potential differs sufficiently
 from the sin-function. They found that if the barrier
between the potential wells is flat enough, the
Peirels-Nabarro (PN) barrier does not decrease monotonically
 with the
coupling constant, as in the ordinary discrete sine-Gordon chain.
It decreases with oscillations, so that the PN barrier experiences
dips,  where it lowers by the order of magnitude. Later \cite{sze00pd}, 
it was found
that if the PN barrier decreases nonmonotonically, there
exist certain velocities, at which even very discrete kinks propagate
with constant shape and velocity. Everywhere in between these
velocities, there exist kinks with oscillatory asymptotics (nanopterons).
Approaching the problem from another side, Schmidt in \cite{s79prb}
has constructed a Klein-Gordon model that allows an exact moving
kink solution of the form $\tanh(n-vt)$ for some specific
value of velocity $v$. Furthermore, Flach and coauthors \cite{fzk99pre}
have shown that for this model the PN barrier is nonzero
(when $v \neq 0$). It was also shown that kinks of the
discrete sine-Gordon equation with  
topological charges greater then one exhibit some
features, similar to those, described above, 
including free propagation at some selected
velocities (see Refs. \cite{pk,bck,sze00pd}).
Note that if we step out from the Klein-Gordon class of
 discrete models (for instance, by introducing anharmonicity
into the interparticle interaction), some new phenomena related
to the kink mobility can appear, but this is out of the
scope of the present paper.
In this context, a few papers (see Refs. \cite{sze97pd,sp} and 
others therein) should also be mentioned. 

Thus, owing to the importance of discreteness
effects in the kink 
dynamics, it would be of big interest to apply these findings
first
for the one-component model of proton transport in HB chains
and then for the two-component model given by the Hamiltonian
(\ref{1})-(\ref{4}). In this context, as found by Duan and Scheiner 
\cite{ds,sd}, a pair of Morse functions,
placed tail-to-tail so as to allow for the approach of the
proton towards the acceptor while it is departing from the
donor (see Fig.~1), provides the best framework 
for reproducing their
potentials obtained from {\it ab initio} calculations.
 It is important that the Morse-type functions
contain parameters with clear physical meaning, which
vary little from one H-bond to the next one in HB systems.

The aim of
this paper is to investigate the properties of the
one-dimensional Klein-Gordon chain with the on-site 
(intrabond) potential
of the double-Morse type. We are going to find both
stationary and dynamic (moving) kink solutions and to show
that kinks can be mobile even if being very narrow.

The paper is organized as follows. In the next section,
we present the Hamiltonian and the equations of motion for
the one-component model. In Sec. III, we study the properties
of the stationary kink solutions. In Sec. IV, the Peierls-Nabarro
potential for the kinks is investigated. Section V is devoted to
the studies of kink mobility. Conclusions are given in Sec.VI.

%
%
\section {The double-Morse proton potential}

In the limit, when the heavy ions are fixed (immobile) at a same
distance $l$ forming a uniform lattice, 
we deal only with the first part
of the Hamiltonian (\ref{2}), where $q_n = u_n$. 
In what follows we adopt the dimensionless description,
where for the dimensionless proton displacement $u_n/l$ we keep
the same notation $u_n$ and use the time unit 
$t_0= l/\sqrt{\varepsilon_0/m_p}$. In these
dimensionless variables, the Hamiltonian (\ref{2}) reads
\begin{equation}
{\cal H}=\; \sum_n \; \bigg[
 \frac {1} {2} \dot{u}_n ^2 
 + \frac{\kappa}{2} (u_{n+1}-u_n)^2 + V(u_n) \bigg] .
\label{5}
\end{equation}
Here and in what follows, the overdot denotes the differentiation 
with respect to $\tau =t/t_0$ and
$\kappa=K_p l^2/\varepsilon_0$ is the dimensionless proton-proton
coupling constant.

As described in Introduction and illustrated by Fig.~1, 
the intrabond proton potential
$V(u)$
can be formed as a result of superposition of two pair
ion-proton interaction
potentials placed tail-to-tail. 
According to the {\it ab initio} studies of Duan
and Scheiner \cite{ds,sd}, these potentials are preferred to be
 chosen of the Morse type.
As a result, the potential $V(u)$ is a symmetric
double-well function
with minima at $u=\pm a $ and 
a maximum at  $u=0$ \cite{zps,zps91,gsz}:
\begin{equation}
V(u)=\bigg[ \;\frac {\alpha
- \cosh{(\beta u)}}{\alpha -1} \; \bigg ] ^2, \; ~
\alpha=\cosh(\beta a) .
\label{6}
\end{equation}
The inequality $\alpha > 1$ ensures the double-well form 
of the function (\ref{6}).  Throughout the paper we take $a=0.25$.
The potential (\ref{6})
is normalized so that the barrier height always equals unity. 
Its shape for different values of
$\beta$ is shown in Fig.~\ref{potential}. As can be seen 
from this figure, the parameter $\beta$ determines the
curvature/flatness of the barrier and its  shape 
strongly depends on this parameter. We assume its values 
to range  over the whole half-axis $0 < \beta < \infty$.
For small $\beta$, the barrier is rather narrow being the limiting
case of the $\phi^4$ model, i.e.,
\begin{equation}
\lim_{\beta \rightarrow
0}V(u) = \left ( 1- u^2/a^2 \right )^2 .
\label{7}
\end{equation}
Increase of $\beta$
makes the barrier more flat and the wells more narrow, so that
the other limit is
\begin{equation}
\label{8}
\lim_{\beta \rightarrow \infty}V(u) = \left \{
 \begin{array}{cccc}
  \infty, && -\infty < u < -a , \\
  0,  && u= \pm a,\\
  1,  && -a < u < a , \\
  \infty, && a < u < \infty .
 \end{array}
 \right .
\end{equation}

%
%

The corresponding equation of motion is the well known discrete
nonlinear Klein-Gordon equation:
\begin{equation}
\ddot {u}_n\;=\;\kappa \;(u_{n+1}-2u_n+u_{n-1})-V'(u_n),~  
 n = 0, \pm 1 , \ldots .
\label{9}
\end{equation}
Here and in what follows
 the prime denotes differentiation of a function with respect to
its argument. 

Before embarking on a complete analysis of the discrete 
equation (\ref{9}), we calculate the dispersion law 
of small-amplitude waves
around one of the two ground states.
As a result, this law is given by the
following equation:
\begin{eqnarray}
\nonumber
\omega^{2}(q)&=&\omega_{0}^2+2 \kappa (1-\cos {q}),\;\\
\omega_0&=& \sqrt{2 \frac{\alpha +1}{\alpha-1}}\beta=
\frac{\sqrt{2}\beta}{\tanh {(\beta a/2)}}.
\label{10}
\end{eqnarray}
The gap of the spectrum depends on the parameter $\beta$;
 for large values of $\beta$, it increases linearly with $\beta$.

%
%
\section {Kink states, their stability and bifurcations}

In this section, we start from the classification of possible stationary
(anti)kink states. To compute these states, we have used both
the conjugate gradients method
for minimization of the stationary part (${\dot u}_n \equiv 0$) in
the Hamiltonian (\ref{5})
and the Newton iteration method for solving the time independent
nonlinear set of equations that originates from Eqs.~(\ref{9}):
\begin{equation}
\kappa (u_{n+1}-2u_n+u_{n-1})=V'(u_n) ,~  n = 0, \pm 1 , \ldots .
\label{11}
\end{equation}
These equations can be rewritten as a two-dimensional map:
\begin{eqnarray}
\nonumber
p_{n+1}&=&u_n, \\
u_{n+1}&=& \kappa^{-1}V'(u_n)+2u_n-p_n,\; n =0, 1, \ldots .
\label{12}
\end{eqnarray}
In general,  maps of this type are chaotic. However, an
on-site potential, for which the map (\ref{12}) is integrable,
 has been found in Ref.~\cite{jbp}.
 The previous knowledge 
on the kink solutions in the
most popular cases of
the discrete nonlinear Klein-Gordon
lattices such as the 
 sine-Gordon or $\phi^4$ chains implies the existence of 
 only two stationary kink states. These states
possess the inversion symmetry with respect to the center of the
kink, being monotonic functions on the lattice.
They connect two hyperbolic
fixed points of the map $(-a,-a)$ 
and $(a,a)$, which are the
ground states of the chain.
 
Adapted to our case of a HB chain with the numbering of ions
and protons according to the sequence 
$\{ \ldots , Q_{n-1}, q_{n-1}, Q_n , q_n , Q_{n+1}, q_{n+1} ,
\ldots \}$,  one
of these stationary states, which has its center positioned at 
a heavy ion (call it an ion-centered kink/antikink),
 say with a number $n_0$, is dynamically {\it stable}, whereas the
 other, with its center positioned in the middle of a 
$n_0$th H-bond, i.e., in between the $n_0$th and the $(n_0+1)$th
ions (call it
a bond- or proton-centered kink/antikink), is 
dynamically {\it unstable}.
 The symmetry of the ion-centered (on the $n_0$th ion, in between 
H-bonds  $n_0 -1$ and $n_0$)
kink/antikink  is defined by the relations
\begin{equation}
u_{n_0-n}=-u_{n_0+n-1}, ~ n =0, \pm 1, \ldots ,
\label{13}
\end{equation}
whereas for the proton-centered
(in the middle of the $n_0$th H-bond, 
in between ions $n_0$ and $n_0+1$) kink/antikink, the
symmetry relation is given by
\begin{equation}
u_{n_0-n}=-u_{n_0+n} , ~ n =0, \pm 1, \ldots  .
\label{14}
\end{equation}
The solutions of these types certainly exist also in our model,
as illustrated by Fig.~\ref{profiles1}.  However, their stability 
properties
appear to be much more complicated and they
are discussed below.
%
%

In general, a deformation of the barrier shape in the potential $V(u)$
leads to a more rich family of kink/antikink solutions. This has been
observed in the previous studies \cite{pr82prb,sze00pd}.
The first feature caused by the deformation of the proton
 potential (\ref{6}) with increase of $\beta$ is
the phenomenon of {\it stability switching}, according to which 
 the two
types of kink solutions with inversion symmetry
defined by Eqs.~(\ref{13}) and (\ref{14})  switch their
stability, while varying the system parameters.
The  second one is the  
appearance of new types of kink solutions. 
For the first of these types 
the symmetry relations (\ref{13}) and (\ref{14}) are not
valid anymore, whereas the other one is 
symmetric with a 
zigzag-like profile at their center, but still has monotonic 
asymptotics as $|n| \rightarrow \infty$.

\subsection{An exactly solvable limit}

To understand better the effect of stability switching,
it is instructive to consider the limit  $\beta \rightarrow \infty$, 
resulting in a similar potential behavior as studied in Ref.~\cite{sze00pd}.
In this limit, the double-Morse potential takes the form 
[see Eq.~(\ref{8})], 
for which the system of equations of motion (\ref{11}) becomes exactly 
solvable because the particles (protons) can appear only either
 in the wells or on the flat region of the proton potential $V(u)$.
There exists an infinite, but a
countable set of the stationary kink solutions with an arbitrary 
number of particles,  $m=0, 1, \ldots $,
lying on the barrier. This number uniquely defines a
kink/antikink solution. The set of these solutions can be written as
\begin{equation}
\label{15}
u_n= \left \{
 \begin{array}{ccc}
  -a, && -\infty < n \le  n_0-\frac{m}{2} -1 , \\
  2a\frac{n-n_0+ 1/2}{m+1}, && n_0-\frac{m}{2} -1 < 
 n < n_0+\frac{m}{2} ,\\
  a,  &&  n_0+\frac{m}{2}  \le n < \infty ,
 \end{array}
 \right .
\end{equation}
for the kink centered on the $n_0$th ion ($m = 0, 2, \ldots $) and 
\begin{equation}
\label{16}
u_n= \left \{
 \begin{array}{ccc}
  -a, && -\infty < n \le n_0-\frac{m+1}{2}, \\
  2a\frac{n-n_0}{m+1}, && n_0-\frac{m+1}{2} < n < n_0+\frac{m+1}{2},\\
  a,  &&  n_0+\frac{m+1}{2}  \le n < \infty ,
 \end{array}
 \right .
\end{equation}
for the kink centered on the  $n_0$th H-bond  ($m =1, 3, \ldots $).
Similarly, the
analytical expressions for the antikink solutions can be obtained.
These stationary solutions can be found, using either energy
arguments based on the Hamiltonian (\ref{5}) or directly from
the equations of motion (\ref{11}), where 
\begin{equation}
\label{17}
\lim_{\beta \rightarrow \infty}V'(u) = \left \{
 \begin{array}{ccc}
 -  \infty, && -\infty < u < -a , \\
  0 ,  && -a  \le  u \le  a , \\
  \infty, && a < u < \infty .
 \end{array}
 \right .
\end{equation}

The energy of both these  (anti)kink solutions  in the limit 
$\beta \rightarrow \infty$  is easily 
calculated and for any integer $m=0, 1, \ldots , $  it reads
\begin{equation}
E_m=E_m(\kappa)=m+  2 \kappa a^2/ (m+1) .
\label{18}
\end{equation}
As illustrated by Fig.~\ref{betainf},
where the linear dependences of the energy $E_m$ on $\kappa$ 
are plotted for different $m$'s that the crossings of these 
dependencies occur at some values of the coupling parameter 
$\kappa$.

%
%

The most interesting points in Fig.~\ref{betainf}  are the 
crossings for the states with
the energies that correspond to adjacent $m$'s, i.e., $m=0$ and 1,
$m=1$ and 2, and so on because they occur at lowest energies. 
Thus, the crossings  when the kink state with $m$ particles on 
the barrier is transformed into the state with  $m+1 $ particles
on it, occur
at the following values of $\kappa$:
\begin{equation}
\kappa_{m+1}^{(c)}  = (m+1)(m+2)/2a^2   , ~
m=0, 1,   \ldots .
\label{19}
\end{equation}
Therefore, depending on the strength of the proton-proton coupling
$\kappa$, the
proton-centered or the ion-centered kink can reach a 
global minimum of the energy $E_m >0 
$ (except for  the ground states, when $E=0$).
In addition,
 it is interesting to notice that at the values of $\kappa$ given 
by Eqs.~(\ref{19}), the inter-bond and the intrabond energies
are equal exactly each other.

\subsection{Numerical results for finite $\beta$'s}

Now let us investigate 
 how the properties of the kink solutions found in the
exactly solvable limit $\beta \rightarrow \infty$ 
 change,  when $\beta$ take finite
values. Thus, changing 
$\beta$ allows us to explore the whole
set of scenarios,  starting from the $\phi^4$ limit and
finishing with the  limit $\beta \rightarrow \infty$.
We start to compute the kink solutions from the anticontinuous limit
($\kappa=0)$, taking the solutions of the exactly solvable limit as
an
initial guess for the Newton iteration method. Then, we increase
$\kappa$ and check how the kink profiles behave.
In Fig.~\ref{energy1},
the energy dependence on the coupling parameter $\kappa$ is plotted
for $\beta=10$. First, let us focus on the behavior of the
solutions with the lowest energies (namely those which correspond to
$m=0$ and $m=1$  in the exactly solvable limit).

%
%

In Fig.~\ref{energy1}, curve 1 corresponds to the ion-centered kink with $m=0$,
the symmetry of which is given by Eqs.~(\ref{13}),  whereas
 curve 2 to the  proton-centered
kink with $m=1$, the symmetry 
of which is given by Eqs.~(\ref{14}). 
Contrary to the previous knowledge  on the  stability properties of
the kink solutions for  the discrete
sine-Gordon and $\phi^4$ models, where the (anti)kinks of 
the symmetry  (\ref{13}) are
always dynamically  stable, while the (anti)kinks of the 
symmetry (\ref{14}) are always
dynamically unstable,
an
interchange of stability is observed for the proton potential
(\ref{6}) with finite $\beta$'s as $\kappa$ varies.
Thus, one can  see that at a certain value of the coupling parameter,
$\kappa \simeq  24.5$,
the energies of the both types of symmetric kinks coincide and
after passing this critical point, the proton-centered
kink appears to be stable, while the ion-centered kink is unstable.
For higher $\kappa$'s, several more interchanges of stability
 take place, with
the energy difference between the sequential kink states that
 decreases with the
growth of $\kappa$.
These transitions of stability take place smoothly all the way 
up to the
continuum limit. We refer to these transitions
as to {\it stability switchings}.
Thus, the solutions with $m>1$, which were clearly separated
from the $m=0$ and the $m=1$ solutions in the exactly 
solvable limit $\beta \rightarrow \infty$, appear
to be smoothly connected with them. In other words, while
the
coupling  $\kappa$ increases, the particles slowly ``climb'' on the
barrier,  so that there is no abrupt transition from the state
with
$m=0$ to the state with $m=2$ or, further, to the states with 
$m=4, 6, \ldots $.
The same can be concluded about the kinks with $m$ being odd. In this
respect, the system still shows similarity with the $\phi^4$ model.

If we focus more carefully on the behavior of the
 system in the vicinity of the
points, where the energies of proton-centered and ion-centered kinks
become equal, we find that a new type of kinks appears. These kinks
 shown in Fig.~\ref{profiles3}
do not exhibit any of the symmetries defined by Eqs.~(\ref{13}) and 
(\ref{14}), but they are doubly degenerate 
related to each other by the inversion
with respect to the crossing point between the line $u_n=0$ and
the line connecting two central particles of the kink [$n=48$
and $n=49$  in Fig.~\ref{profiles3}(a)].
The energy of these kink states with broken symmetry 
is always larger than  the energy of the symmetric
(ion-centered and proton-centered) kinks. The asymmetric (anti)kinks 
are always linearly
unstable. In Fig.~\ref{energy1},  
their energy is shown by curve 3.
%
%

The kinks of the last type are shown in Fig.~\ref{energy1} 
by curves 4 and 5. This is what
happens to the solutions with $m>1$ obtained in the 
limit $\beta \rightarrow \infty$, when we are moving from 
the anticontinuous limit ($\kappa=0$). 
Instead of
attaining a regular monotonic form, the kink profiles
with several particles on the barrier develop a zigzag-like 
structure in their centers, as demonstrated by Fig.~\ref{profiles2}.
For finite $\beta$'s the zigzag-like solutions are
linearly unstable and therefore we do not study 
them here in more detail. 
%
%

Instead, we focus on the behavior of the {\it monotonic} 
kink solutions with increase
of the coupling $\kappa$  in the vicinity of the stability switchings.
To understand this effect better, in Fig.~\ref{bifurcation}, we have
plotted the position
of the $(N/2)$th proton ($N$ is the total number of 
H-bonds in the
chain) as a function of the
coupling $\kappa$. Here the sequence of pitchfork bifurcations
is clearly seen. Curve 1 corresponds to the kink centered
in the middle of  the $(N/2)$th bond, i.e., when the $(N/2)$th proton is
always a central particle of the kink with $u_{N/2} \equiv 0$. 
The displacements of the $(N/2)$th proton for the ion-centered kinks 
positioned
on the $(N/2 +1)$th and the $(N/2)$th ions are shown by 
curves 2 and 3, respectively. When increasing $\kappa$, the $(N/2)$th
proton moves slowly out of the well. At a certain value
of $\kappa$, more specifically, at $\kappa \simeq 22.5$, the  pitchfork
bifurcation of the proton-centered kink takes place. This configuration
retains its stability and two new solutions (both linearly unstable)
appear. These are precisely those asymmetric kinks
(see curves 4 and 5 in Fig.~\ref{bifurcation}), with their shape shown in
Fig.~\ref{profiles3}. At the beginning, they look like slightly
distorted proton-centered kinks, but with the growth of $\kappa$,
they change more and more towards the ion-centered configuration.
Eventually, the second pitchfork bifurcation takes place
at $\kappa \simeq 26.2$.  The asymmetric
kinks join the ion-centered kinks (junction of curves 3 and 4,
and curves 2 and 5) and the ion-centered configuration 
loses its stability.
%
%
Now one can clearly see that two identical kinks, shifted by one
lattice spacing with respect to each other, are connected via such
a bifurcation sequence.
Thus, this pitchfork bifurcation is nothing  but a transition of the
kink from one position to another position, 
one lattice period forward or
backwards.
This cascade of bifurcations can be continued further up or down in
$u_n$'s
or, in other words, two or more sites backwards or forward. Similar
bifurcation scenario for discrete breathers in the ac-driven
and damped Klein-Gordon lattice has been reported in \cite{mfmf01pre}.

For higher values of $\beta$, the effect of stability switchings
exists, being more pronounced because the switchings start at smaller
$\kappa$ and take place more frequently (see Fig.~\ref{energy2}).
Another feature that appears from the exactly solvable limit 
$\beta \rightarrow \infty$ is as follows.
Curve 1 of  Fig.~\ref{energy2} corresponds to the ion-centered
kink and curve 2 to the proton-centered kink. They
cross each other at $\kappa \simeq 17.4$, where the 
ion-centered
kink loses its stability,  disappearing shortly
 at $\kappa \simeq 28.6$. Here the same pitchfork bifurcation
scenario takes place, but the distance between the first and
the
second bifurcations is much larger than in the $\beta=10$ case.
In the meanwhile, a bit earlier,
at $\kappa \simeq 23.4$, a new family of ion-centered
kinks appears (curve 4). This curve corresponds to the
ion-centered kinks with two protons on the barrier. Thus,
one can observe the coexistence of two different kink solutions 
with the
same type of 
symmetry (for more details see the upper inset of
Fig.~\ref{energy2}). This coexistence takes place on a rather narrow
interval of $\kappa$ and one of the coexisting kinks
is unstable, but still this phenomenon clearly originates
from the limit $\beta \rightarrow \infty$.
%
%
In Fig.~\ref{fig10}, we show the shape of two coexisting kinks.
One of them (more narrow) corresponds to curve 1 of
Fig.~\ref{energy2} and the second one, which is
more broad, corresponds to curve 4 of this figure.
These kinks  have zero and two protons
on the barrier, respectively.
When $\kappa$  increases further, the stability switchings occur
between curves 2 and 4 under the same scenario as before
for $\beta=10$ (see for details the lower inset of Fig.~\ref{energy2}).
Zigzag-like kinks are also presented in this case, as
shown by curves from 6 to 8.
%
%

When increasing $\beta$ even more, the coexistence of different
kinks with the same symmetry is even more pronounced. We have checked
the case of $\beta=50$ and discovered for this value that several cases
of this coexistence for different kinks take place and they are much more
pronounced. Thus, this case is rather close
to the exactly solvable limit
$\beta \rightarrow \infty$.

\subsection{Analytical approximation for finding kink
solutions}

One can use a simplification of the potential (\ref{6}) in order
to obtain analytically an exact  (anti)kink solution. To this end, 
we approximate
both the 
wells of the potential by parabolas connected by the constant equal to
 the barrier height as follows:
\begin{equation}
V(u) \simeq  \left \{
 \begin{array}{ccc}
  (\omega_0^2/2)(u+a)^2,  & & -\infty <  u   <  -b , \\
  1,   & & -b \le    u   \le  b ,   \\
  (\omega_0^2/2)(u-a)^2, & &   b <  u   <  \infty ,
 \end{array}
 \right. 
\label{20}
\end{equation}
where $ b= a-  \sqrt{2}/ \omega_0$. 
Schematic description of this ``parabola-constant'' 
approximation is presented in Fig.~\ref{approx} by 
thick solid lines. The thin solid lines
show the original potential  (\ref{6})
 with $\beta=20$. The approximation
is expected to work well, when the barrier is flat enough, i.e., 
when $\omega_0 \gg \sqrt{2}/a $. Within this approximation,
we are able to solve the problem of finding stationary kink solutions
analytically.
%
%

Let  $m=0, 1, \ldots ,$  be the number of protons 
on the barrier of the potential
(\ref{20}).  Then the discrete kink profiles are  given by 
\begin{equation}
\label{21}
u_n= \left \{
 \begin{array}{cccccc}
  -a + A_m e^{ \lambda ( n-n_0 + m/2 +1 )}   \\ 
 \mbox{if}~ -\infty < n \le n_0- m/2  -1 , \\
  (n-n_0+ 1/2 )D_m  \\
\mbox{if} ~  n_0-m/2 -1  < n < n_0+ m/2  , \\
  a  -  A_m e^{ - \lambda ( n-n_0 - m/2 )}  \\  
\mbox{if}~  n_0+  m/2  \le n < \infty ,
 \end{array}
 \right .
\end{equation}
for the kink centered on the $n_0$th ion ($m =0, 2 , \ldots $) and
\begin{equation}
\label{22}
u_n= \left \{
 \begin{array}{cccccc}
  -a   + A_m e^{\lambda [ n-n_0 + (m+1)/2 ]}   \\
  \mbox{if}  ~ -\infty < n \le n_0-(m+1)/2 , \\
  (n-n_0 ) D_m   \\
   \mbox{if} ~ n_0-(m+1)/2 < n < n_0+ (m+1)/2 ,  \\
  a -  A_m e^{ - \lambda [ n-n_0 - (m+1)/2 ] }  \\
  \mbox{if}  ~  n_0+ (m+1)/2  \le n < \infty ,
 \end{array}
 \right .
\end{equation}
for the kink solution  centered on the  $n_0$th H-bond  
($m= 1, 3 , \ldots $).  Here 
$\lambda$ is a ``localization'' parameter that measures the transition
width in the (anti)kink profile  between 
the uniform distribution of protons on the barrier and the 
(anti)kink asymptotics $u_n \rightarrow \pm a$. It is given by a positive
root of the equation 
\begin{equation}
\cosh \lambda = 1 + \omega_0^2 /2\kappa .
\label{23}
\end{equation}
The other two parameters, the amplitude $A$
and the uniform distance between the nearest-neighbor protons on the flat $D$
can be expressed through the  localization  parameter $\lambda$ as
\begin{eqnarray}
A_m  & =& {2a \over (m+1)e^\lambda - m+1 } , \nonumber \\
D_m  & =& {2a e^\lambda  \over (m+1) e^\lambda - m+1 } .
\label{24}
\end{eqnarray} 
As follows from Eqs.~(\ref{10}) and (\ref{23}), 
the limit $\lambda \rightarrow \infty$ (when $\omega_0^2/\kappa 
\rightarrow \infty$) is more general than the limit 
$\beta \rightarrow \infty$ because $\lambda$ contains both
$\beta$ and $\kappa$. Therefore one can check that
Eqs.~(\ref{21})-(\ref{24}) are reduced
to the stationary kink solution given by Eqs.~(\ref{15}) and (\ref{16})
as $\lambda \rightarrow \infty$. In particular, the amplitude $A_m$ 
and distance between the protons on the barrier $D_m$ tend to 
zero and $2a/(m+1)$, respectively.

Using Eqs. (\ref{21})-(\ref{24}) one can easily compute the  energy of 
both the kink configurations:
\begin{equation}
E_m = m + 2\kappa a^2 {\tanh(\lambda/2) \over 1 +m \tanh(\lambda/2)} . 
\label{25}
\end{equation}
Similarly, this expression is also transformed to the energy (\ref{18})
as $\lambda \rightarrow \infty$.

Now  we investigate the behavior of the energy difference 
\begin{equation}
\Delta E_{m+1}  (\kappa, \beta)=E_{m+1}-E_m , ~m=0, 1,  \ldots  .
\label{26}
\end{equation}
We find that this difference as a function of $\kappa$ has a number of 
zeroes
and these zeroes depend on the parameter $\beta$.
Thus, within our approximation one can predict the effect of
switching of the stable and unstable kink configurations.
In Table \ref{tab1},  we show the values of the coupling
parameter $\kappa$, for which the first [i.e., when 
$\kappa = \kappa_1$ and  $m=0$; see also Eq.~(\ref{19})] 
switching of the kink
stability takes place.
%
\begin{table}
\caption{
Comparison of numerically and
analytically calculated values of $\kappa$ for which
the first
($m=0$) stability switching occurs.}
\label{tab1}
\begin{tabular}{c|c|c}
$\beta$  & $\kappa_1 $, numerical   & $\kappa_1$, analytical     \\
\hline
5         &   64.10  &   18.0   \\
10        &   24.547 &   17.069 \\
20        &   17.383 &   16.329 \\
50        &   16.172 &   16.052 \\
$\infty $ &   16.0   &   16.0   \\
\end{tabular}
\end{table}
We see that the approximation works fairly well when $\beta$ is rather
large and the barrier between the wells is close to being completely
flat. It is improving with increase of $\beta$ and in the
limit $\beta \rightarrow \infty$, our approximation coincides with the
exact result, shown above. Thus, switchings of the stability of
kink states with different symmetries is a generic effect
that does not depend on a specific model,  but on the properties of the
on-site potential.
Another piecewise approximation of the 
proton potential $V(u)$ for a HB chain 
has been constructed earlier by Weiner and Askar \cite{wa},
using alternatively inversed parabolas. Our
potential approximation (\ref{20}) seems to be more appropriate
for the  studies in detail 
of flatness effects of the on-site (intrabond) potential
and more close to the realistic double-Morse potential (\ref{6})
if $\beta$ is not so small.

Note that the zigzag-like kink profiles obtained above numerically
shown, e.g., in Fig.~7
can also be given analytically within the approximation (\ref{20}).
Indeed, the ion-centered kink shown in Fig.~7(a) is described by
\begin{equation}
\label{27}
u_n= \left \{
 \begin{array}{ccc}
  -a   + B_0 e^{\lambda ( n + 2)}, ~n=-2, -3, \ldots , \\
  u_0 = -u_{-1} = \xi_0 , \\
     a -  B_0 e^{ - \lambda ( n-1)} , ~n=1, 2, \ldots ,
 \end{array}
 \right .
\end{equation}
whereas the proton-centered profile illustrated by Fig.~7(b) is 
given by
\begin{equation}
\label{28}
u_n= \left \{
 \begin{array}{ccc}
  -a   + B_1 e^{\lambda ( n + 2)} , ~n=-2, -3, \ldots ,  \\
  u_0 = 0, ~u_1= -u_{-1} =  \xi_1 ,  \\
     a -  B_1 e^{ - \lambda ( n-1)}  , ~n=1, 2, \ldots ,
 \end{array}
 \right .
\end{equation}
where $\lambda$ is given by Eq.~(\ref{23}) and 
\begin{eqnarray}
B_m & =& 2a {e^\lambda +  e^{-\lambda} -m/2 -1 \over
e^\lambda \left( e^\lambda -m +1 \right) } , \nonumber \\
\xi_m &= & a { 3 - e^\lambda -2 e^{-\lambda} \over 
e^\lambda -m +1 } , ~~m=0, 1 .
\label{29}
\end{eqnarray}
Using the last equations, the kink energy can be calculated in a 
similar 
way as the energy (\ref{18}). In the limit $\lambda 
\rightarrow \infty$, the energy of the zigzag-like kinks, profiles 
of which
are  illustrated by Fig.~7, become
\begin{equation}
E_0 = 6\kappa a^2 ~~\mbox{and}~~ E_1= 1+ 5\kappa a^2 .
\label{30}
\end{equation}
These expressions clearly show that for this type of kinks
their energies also can coincide for certain values of
coupling $\kappa$ if $\beta$ (and, consequently, $\lambda$) is
large enough.

%
\subsection{Elementary excitations on the kink background}
%
%

Let  $u_n^{(0)}$  be a stationary kink solution of Eq.~(\ref{11}), i.e., 
 a fixed point of the map (\ref{12}). We are interested in the properties
of small-amplitude excitations on the kink background. Linearizing
Eq.~(\ref{9}) around the stationary kink solution according to 
\begin{equation}
u_{n}=u_{n}^{(0)}+A_n e^{i \Omega \tau}  ,
\label{31}
\end{equation}
we arrive at the eigenvalue problem
\begin{equation}
\hat L {\bf A} = \Lambda  {\bf A}, \;{\bf A}=\{\ldots , A_{n-1}, A_n, A_{n+1} ,\ldots , \} .
\label{32}
\end{equation}
Here the operator $\hat L$ [the Hessian of the Hamiltonian
Eq.~(\ref {5})]  acts on a vector ${\bf
 A}$
as
\begin{equation}
 (\hat L {\bf A} )_n  =
 -\kappa(A_{n+1}-2A_n+A_{n-1})+V_nA_n   ,
\label{33}
\end{equation}
where $V_n  = V'' [u_n^{(0)}]$.  The operator $\hat L$
 is a symmetric (so all eigenvalues are real) tridiagonal matrix
and the spectral parameter is $\Lambda\equiv\Omega^2$.
This eigenvalue problem can be treated as a quantum-mechanical problem
of a particle, trapped in a single-well spatially
discrete potential formed by the kink. Its depth depends on the
curvature/flatness
 of the proton potential in the middle of the H-bond, $V''(0)$,
and tends to $\omega_0^2$ as $n \rightarrow \pm \infty$.

The eigenvalues of the problem also give information about the
linear
stability of the kink solution. If there exists at least one
eigenvalue  $\Lambda=\Omega^2<0$, the linear excitation on the
kink grows exponentially in time and the corresponding kink solution
is linearly unstable. Otherwise,  it is linearly stable.
The stability of stationary kink solutions is determined by the system 
parameters (in
our case, by the
curvature/flatness parameter  $\beta$ and the coupling $\kappa$).

In Fig.~\ref{eigenvals},  we depict the dependence of the eigenfrequencies
 $\Omega_n$'s
on the coupling parameter $\kappa$ for different curvatures $\beta$.
The spectrum consists of two parts. The first one describes the
eigenfrequencies that lie above the potential
$V_n$  and therefore determine delocalized eigenvectors corresponding to
the linear (phonon) spectrum of the lattice. The second part consists
of eigenfrequencies that lie below $\omega_0^2$ corresponding to spatially
localized eigenvectors. The lowest eigenfrequency is the Goldstone
mode, which is universally presented in all the Klein-Gordon models.
%
%


The second mode corresponds to small-amplitude 
oscillations of the kink
core, being often referred to as the Rice mode. In the continuum 
$\phi^4$ model, there are
only two localized modes: the Goldstone mode and the Rice mode.
The properties of the internal modes can be significantly
altered due to different factors such as change of the nature
of the interparticle interaction \cite{mgms} or change
of the shape of the on-site potential \cite{bkp97}.
If $\beta$
is not so large, the behavior of the eigenfrequencies and
eigenvectors (see Fig.~\ref{eigenvectors}) of our system is
reminiscent to that of the $\phi^{4}$ model,
 as shown in panels (a) and (b) of
Fig.~\ref{eigenvals}. However, several differences occur. Thus,
 the Goldstone mode collides with the zero axis and becomes
unstable [see Fig.~\ref{eigenvals}(a)] for the ion-centered kink.
Meanwhile, for the proton-centered kink [see Fig.~\ref{eigenvals}(b)], 
the 
Goldstone mode was initially unstable and became stable later on.
The stability switchings are caused by the pitchfork bifurcations
as described at the beginning of this section. Another
difference
is the appearance of the new eigenfrequency for the proton-centered kink
with
the eigenvector which has two nodes [see Fig.~\ref{eigenvectors}(e)].

%
%
Thus, one can see
that curvatures play an important role in the properties of the
eigenfrequencies of our system. Increasing $\beta$ further, we
observe more pitchfork bifurcations of the Goldstone mode
and the appearance of more localized modes. In panels (c) and (d), we show
how the eigenfrequencies behave for $\beta=10$. We observe more
localized internal modes, some of them surviving in the continuum
limit and some of them disappearing there. The shape of the corresponding
eigenvectors (see Fig.~\ref{evecs2})
behaves accordingly to the
 wavefunction shape of the bound states
of the quantum-mechanical Schr\"{o}dinger equation.
%
%

%
%
\section {Properties of the Peierls-Nabarro barrier}

 In general, the propagation of 
 topological solitons (kinks and antikinks) in lattices
are subject to
 their discreteness.  The discreteness effects can be described by
a  spatially periodic potential with the period coinciding with the lattice
spacing $l$, known as a Peierls-Nabarro (PN) potential.
 The PN potential
$E_{PN}(n_c)$ is a function of the kink center
position
\begin{equation}
n_c=\sum_n n \frac{u_{n+1}-u_{n-1}}{2(u_{\infty}-u_{-\infty})}=
\frac{1}{4a}\sum_n n ({u_{n+1}-u_{n-1}}) ,
\label{34}
\end{equation}
and the height of the PN barrier equals
\begin{equation}
\delta E=E(n_{c, max}) - E(n_{c, min})  .
\label{35}
\end{equation}
This difference measures the activation energy for the kink propagation
though one lattice spacing. Here $n_{c,max}$ and $n_{c,min}$
are the positions of the kink, where the PN potential has
its maxima or minima, respectively. The minimum position
 $n_{c,min}$ coincides with the kink position, which is
a
global minimum of the kink energy. Thus, normally $n_{c,min}$ is
an integer or a half-integer.

It is a well established fact that for the $\phi^4$ and sine-Gordon
models, the height of the PN barrier coincides with the difference
between the energies of the proton-centered and ion-centered
kinks.  As demonstrated above, the deformation of the
 proton (intrabond) potential
$V(u)$
leads to the switching of stability of these two states. This
does not mean, however, that the PN barrier disappears, when the
energy of these states coincides
because
 the site-centered and proton-centered states do not represent the
states with the highest and the lowest energy of the PN potential
 \cite{pr82prb,sze00pd,bkp97}.

In  Fig.~\ref{PN}, some  examples of variation of the energy $E(n_c)$ for
different  values of $\kappa$ are presented.
Panel (a) corresponds to the situation, when the proton-centered
kink is an energy minimum and the ion-centered one reaches a
maximum. The second panel (b) demonstrates the case, when the
first pitchfork bifurcation occurs  and the asymmetric
kink appears. Panel (c) corresponds to the case, when
the stability switching happens and both the proton- and
ion-centered kink states get minima,  and the asymmetric kinks
are maxima. The period of the PN potential is decreased by one half.
After that, as shown in panel (d), the proton-centered kink
becomes unstable and the ion-centered one stable. The last
panel (e) is obtained after the second pitchfork bifurcation has
occured and the asymmetric kinks disappear.
%
%

Dependence of the PN barrier is non-monotonic not only in
$\kappa$ but also on $\beta$. In Fig.~\ref{PNb}, we show
the dependence of the PN barrier on the parameter $\beta$
for a fixed value of $\kappa$. We observe a minimum at
$\beta \simeq 8.3$, which is obviously caused by the
stability switching. Indeed, as shown by the upper inset,
the minimum of the PN potential moves from a
half-integer value to an integer one. Thus, the
stable kink configuration changes from the  ion-centered
 to the proton-centered kink.
Note, that the height 
of the PN barrier does not attain zero
 (see the lower inset), but decreases by two order of magnitude.
%
%

%
\section{Discrete kink mobility}
%

Another interesting consequence of the deformation of the
 on-site
potential barrier is the possibility of the existence of non-oscillating
travelling kinks. In the continuum Klein-Gordon models which admit 
 moving topological solitons (kinks and antikinks),  the domain of 
admissible  soliton velocities $s$  is the interval $0 \le s < c_0$, 
where $c_0=\sqrt{\kappa}$ is the
characteristic velocity. Thus, kinks in these continuum models
are a one-parametric family of solutions with the kink velocity
$s$
as a parameter.
In a general case of the discrete
Klein-Gordon model, this family is reduced to a {\it discrete}
set of travelling kink solutions with some
velocities $s_0$, $s_1, \ldots ,$ $s_k$, $s_0=0$, $s_k < c_0$. In 
the $\phi^4$ and
the
sine-Gordon models, there exists only $s_0$. 
In general, everywhere in between $s_n$'s,
there exist moving kinks with {\it oscillating} asymptotics
known as  {\it nanopterons}. 
The existence of velocities $s_n \neq 0$
has been shown both numerically \cite{sze00pd} and
analytically \cite{s79prb,fzk99pre} for several models.
This effect is due to the properties of the on-site
(intrabond)
potential $V(u)$ and, more precisely, the flatness of its
barrier which, in its turn, causes the stability switchings
described in the previous sections.

 For finding non-oscillating
kink solutions, we have used
a pseudospectral method \cite{fe}. The method
allows us to find the travelling-wave kink solutions of the type
\begin{equation}
u_n(t)=u(n-s\tau)\equiv u(z) ,
\label{36}
\end{equation}
solving the
 differential equation with advanced and delay terms:
\begin{equation}
s^2u''(z)=\kappa[u(z+1)-2u(z)+u(z-1)]-V'[u(z)] .
\label{37}
\end{equation}
Dependence of the kink velocity on the coupling parameter $\kappa$ is
shown in Fig.~\ref{sk}. As follows from this figure, it starts (see curve 1)
at the
value of $\kappa$, which is close to that when the
pitchfork bifurcation takes place (see Fig.~\ref{bifurcation}). Then
the
velocity grows with $\kappa$. The second velocity dependence 
starts at $\kappa$
being close to the point of the second pitchfork bifurcation.
%

Dependence of the kink velocity on the parameter $\beta$ has similar
behavior, as  demonstrated by Fig.~\ref{sbeta}. It is interesting to
note that the value of the velocity of the moving kink does not
decrease down to zero as $\kappa$ or $\beta$ decreases, but
stops at some finite $s$ instead. This means that the kinks
should have sufficient kinetic energy to overcome the pinning
effects.
%

In Fig.~\ref{fig19}, we plot the profiles $u(z)$  of the
moving kinks that correspond to both two curves of Fig.~\ref{sk}. 
The
profile in panel (a) corresponds to the moving kink from curve 1,
whereas the profile in panel (b) to the kink from curve 2.  In both
the
cases, the coupling constant was the same:  $\kappa=100$. The second
kink appears to be wider and this can be explained by the fact that
its velocity is more than two times smaller.
The insets of the figures show the velocity profiles.
%
%

In Fig.~\ref{20}, we investigate the behavior of moving kinks
for different $\beta$'s. Increase of the barrier flatness 
is reflected in change of the shape at the kink center.
We have considered the kinks corresponding to curve 2 
of Fig.~\ref{sbeta}, when
the
coupling is fixed:  $\kappa=120$. Deformation
of the kink profile can easily be seen on the velocity profile
shown in the inset. The kink profile experiences 
 deformation of its slope part, which is
seen as a dip in the velocity profile.
The dip grows with increase of $\beta$. More flat the barrier becomes,
more possibilities of the kink's profile deformation.
%
%

These results demonstrate that the existence of a finite set of
velocities, at which the kink can move with constant shape is a generic
effect. Its appearance is due to the shape of the on-site 
(intrabond proton) potential.
If the proton potential has a barrier which is flat enough to
allow the symmetry switchings (accompanied by the
pitchfork bifurcations), the Peierls-Nabarro barrier experiences
lowering for specific values of the system parameters. Thus, it becomes
possible that the kinetic energy of the kink is sufficient to
overcome the pinning forces of the lattice.

%
%
\section{Conclusions}

We have studied the dynamics of the one-dimensional Klein-Gordon
lattice with the on-site potential of the double-Morse type. This is a
physically motivated model, which is the simplest one for the 
proton transport in a
hydrogen-bonded chain, where the on-site potential plays the role
of the potential for proton transfers in the hydrogen bond. 
Therefore throughout this paper we call it an {it intrabond} 
proton potential. A Morse-type function was found to offer 
the best combination of accuracy in reproducing 
quantum-mechanically computed potentials \cite{ds,sd}.
The model has two parameters, 
the proton-proton coupling $\kappa$ and the anharmonicity of 
the Morse potential,
 the curvature parameter $\beta$. 
The anharmonicity parameter is responsible for the shape
of the intrabond potential $V(u)$, especially on the convexity of
its barrier. For larger $\beta$, the barrier becomes more flat
and the wells become more narrow. Changing this parameter,
one can explore the variety of possible intrabond potentials,
starting from the $\phi^4$ model (as $\beta \rightarrow 0$) 
and finishing 
with the  exactly solvable limit $\beta \rightarrow \infty$.

Deformation of the
barrier of the intrabond potential becomes crucial for the
properties of the stationary kink solutions. While the $\phi^4$ limit
allows us only {\it two} types of stationary kinks:  ion-centered
and proton-centered with their stability properties being constant
for any coupling $\kappa$, 
the opposite limit shows the existence of an infinite
countable set of stationary kink solutions. For some values of
the
coupling $\kappa$ forming an infinite countable set, the states 
with different symmetries can have the same
energy.
In between these two limiting cases, some of the kink properties 
survive from the exactly solvable limit $\beta \rightarrow \infty$. 
First of all, the
stability properties of the kink solutions depend
drastically on the coupling parameter $\kappa$. With increase
of $\kappa$, the initially stable ion-centered kink becomes
unstable, while the proton-centered kink gets  stable. With
further increase of $\kappa$, these {\it stability switchings}
which are, in fact, pitchfork bifurcations, go on several times.
Another result can be seen for rather high $\beta$, at least, for
$\beta>20$, in our calculations. It is the coexistence of
several kink solutions of the same symmetry for the same
$\kappa$. This is a left-over from the exactly solvable limit
and it disappears with lowering $\beta$.
An
analytical approximation has been constructed to show the
effect of the symmetry switchings analytically and
to confirm that the effect is not confined to a specific
model, but has a universal nature.

The stability switchings contribute to the non-monotonic
behavior of the Peierls-Nabarro (PN) barrier. The barrier decays
with increase of $\kappa$, however, experiences local
minima at the stability switching points, where its value
decreases by one order of magnitude. The same happens
for the dependence of the PN barrier on $\beta$.
This phenomenon, in its turn, assists the kink mobility and
leads to the appearance of a finite set of velocities for which
the
propagation of very {\it narrow} kinks is possible. This is a
generic effect, attributed to the shape of the on-site
(intrabond)
potential, and also is not confined to a specific model. Note
that the PN barrier is not required to vanish completely
(see Ref.~\cite{fzk99pre}). Simply, the barrier is low enough
for the kinetic energy of the kink to carry it over the
barrier.

Thus, we have concluded that highly mobile kinks are possible
in our model of the proton transport in hydrogen-bonded chains
 even for
those proton-proton nearest-neighbor interactions,
 when the proton kinks are very narrow.
This gives us a reason to believe that the soliton mechanism
of
 proton transfers can work for physically reasonable
values of the proton-proton coupling $\kappa$.

This work has benefited from discussions with M.~Peyrard.
We acknowledge financial support from the
 RTN
  Project No. LOCNET HPRN-CT-1999-00163,
 the INTAS Grant No. 97-0368, and the Danish Research Agency.


%
%

\newpage
\newpage

%
%
\begin{figure}[htb]
\caption{
 Schematics of interactions in the hydrogen bond.}
\label{fig1}
\end{figure}

%
%
\begin{figure}[htb]
\caption{
The shape of the intrabond potential $V(u)$ given by 
Eq.~(\ref{6}) with $a=0.25$ for $\beta=5$
(curve 1), $\beta=20$ (curve 2), and $\beta= 50$ (curve 3).}
\label{potential}
\end{figure}

%
%
\begin{figure}[htb]
\caption{Profiles of monotonic symmetric kinks with
 $\beta=5$ and $\kappa=30$: (a)
ion-centered kink and (b) proton-centered kink. }
\label{profiles1}
\end{figure}

%
%
\begin
{figure}[htb]
\caption{Dependence of the kink energy $E_m$, 
$m=0,~1,~2,~3$, and 4, given by Eq.~(\ref{18}),  in the
exactly solvable limit $\beta \rightarrow \infty$ 
on the coupling parameter $\kappa$.}
\label{betainf}
\end{figure}

%
%
\begin{figure}[htb]
\caption{ Dependence of energy for $\beta$=10 on coupling
$\kappa$ for  symmetric ion-centered kink (curve 1), symmetric
 proton-centered
kink (curve 2), and zigzag-like kinks (curves 4 and 5).
The inset shows more detailed behavior in the vicinity
of stability switchings and curve 3 corresponds to the kink with
asymmetric profile.
 Solid lines correspond to stable
states and dashed lines to unstable ones.}
\label{energy1}
\end{figure}

%
%
\begin{figure}[htb]
\caption{Profiles of monotonic asymmetric kinks with
 $\beta=10$ and $\kappa=24$. Inversion of these profiles
is clearly seen from comparison of panels (a) and (b).}
\label{profiles3}
\end{figure}

%
%
\begin{figure}[htb]
\caption{Zigzag-like kink profiles  for $\beta=10$ and
$\kappa=8$:
(a) ion-centered and (b) proton-centered kinks. }
\label{profiles2}
\end{figure}

%
%
\begin{figure}[htb]
\caption{Dependence of the displacement of the central particle 
 from equilibrium (at $u_{N/2} = 0$) 
on the coupling $\kappa$ for proton-centered
 (curve 1), ion-centered (curves 2 and 3), and
asymmetric (curve 3) kinks with $\beta$=10.
Solid lines show stable kinks and dashed lines  unstable ones.}
\label{bifurcation}
\end{figure}

%
%
\begin{figure}[htb]
\caption{ Dependence of the kink energy for $\beta$=20 on the
coupling $\kappa$ (see text for details).
The solid line shows stable
states and the dashed lines unstable ones.}
\label{energy2}
\end{figure}

%
%
\begin{figure}[htb]
\caption{Kink solutions
corresponding to curves 1 ($\circ$) and 4 ($+$)
of  Fig.~\ref{energy2} for $\kappa=25$.}
\label{fig10}
\end{figure}

%
%
\begin{figure}[htb]
\caption{Schematic representation of the approximate
potential (\ref{20}). }
\label{approx}
\end{figure}

%
%
\begin{figure}[htb]
\caption{Dependence of the system eigenfrequencies $\Omega_n$
on the coupling parameter $\kappa$ for  $\beta=5$:  (a) ion-centered
and (b) proton-centered kinks and for $\beta=10$:  (c) ion-centered
and (d) proton-centered kinks. Curves depicted by small dots
correspond to cases, when $\Omega_n$'s  are purely
imaginary and therefore Im $\Omega_n$'s are plotted instead 
(see text for details).}
\label{eigenvals}
\end{figure}

%
%
\begin{figure}[htb]
\caption{Example of eigenvectors for the case with  $\beta=10$ and
$\kappa=10$.
For the ion-centered kink:  (a) the eigenvector of the lowest localized
mode and (b) the eigenvector of the first excited localized mode.
For the proton-centered kink: the
eigenvectors of (c) the lowest, (d) the first, 
and (e) the second localized modes.}
\label{eigenvectors}
\end{figure}

%
%
\begin{figure}[htb]
\caption{Example of eigenvectors for $\beta=10$ and
$\kappa=30$: (a), (b), (c), and (d) - eigenvectors of the localized modes for the
ion-centered kink from the lowest to the highest mode;
(e), (f), and (g) - the same for the proton-centered kink.}
\label{evecs2}
\end{figure}

%
%
\begin{figure}[htb]
\caption{ Peierls-Nabarro potential for $\beta=10$ and
(a) $\kappa=70.5$, (b) $\kappa=71.22$, (c) $\kappa=71.288$, (d)
$\kappa=71.35$, and (e) $\kappa=72$.}
\label{PN}
\end{figure}

%
%
\begin{figure}[htb]
\caption{ Dependence of the height of the PN barrier on the
parameter $\beta$
for $\kappa=30$. The upper inset shows the
shape of the PN barrier before the minimum at $\beta=7.8$
(curve 1) and after the minimum at $\beta=8.8$ (curve 2).
The lower inset shows more detailed behavior of $\delta E$
around its minimum.}
\label{PNb}
\end{figure}

%
\begin{figure}[htb]
\caption{Normalized velocity of the non-oscillating kink motion
against the coupling  parameter $\kappa$ for $\beta=10$.}
\label{sk}
\end{figure}

%
\begin{figure}[htb]
\caption{Normalized velocity of the non-oscillating kink motion
against
parameter $\beta$ for $\kappa=120$.}
\label{sbeta}
\end{figure}

%
%
\begin{figure}[htb]
\caption{Examples of moving antikinks at $\beta=10$ and $\kappa=100$
with velocities: (a) $s=6.227$ and (b) $s=2.463$. Circles show
positions of lattice sites.}
\label{fig19}
\end{figure}

%
%
\begin{figure}[htb]
\caption{Examples of moving antikinks at $\kappa=120$:
(a)  $\beta=12$ and $s=4.154$,  (b) $\beta=15$ and $s=4.551$,
(c) $\beta=20$ and $s=4.734$. Circles show
positions of lattice sites.}
\label{fig20}
\end{figure}

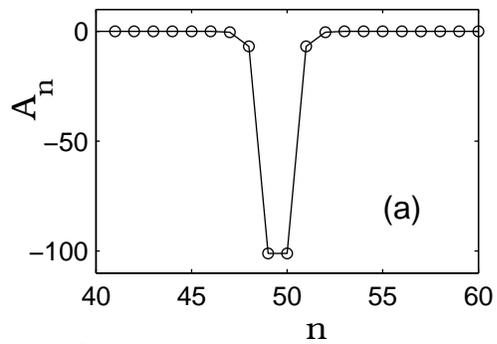
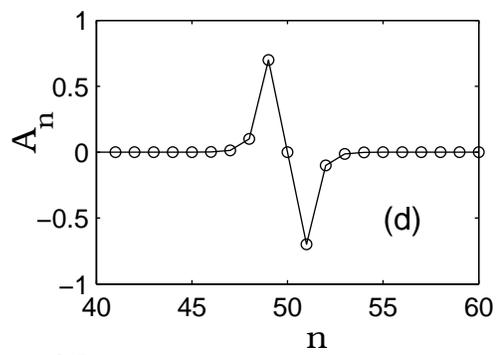
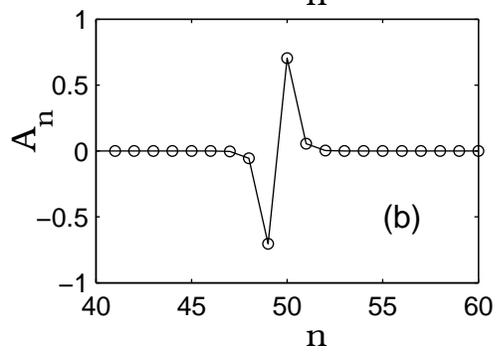
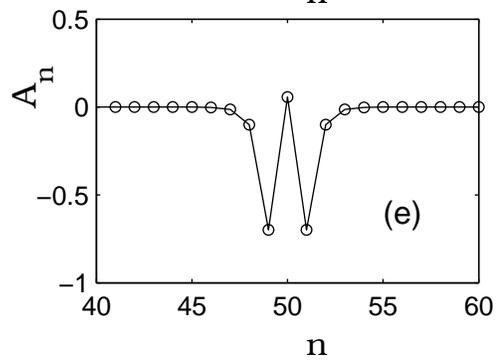
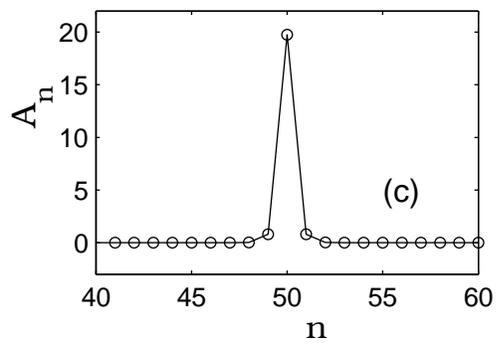
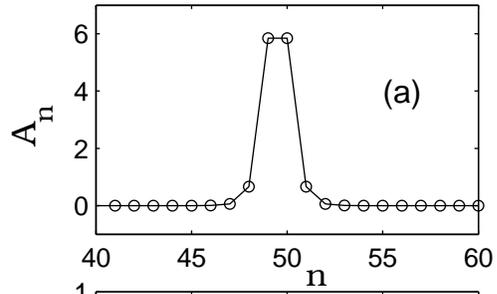
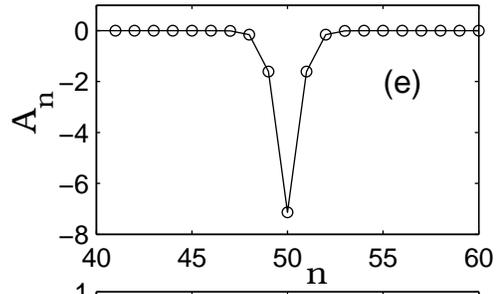
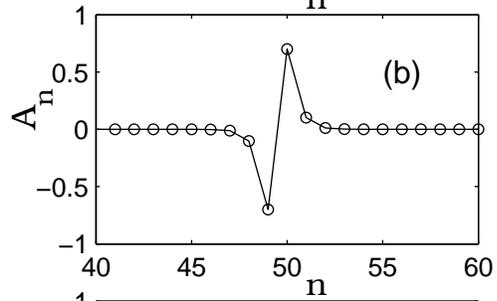
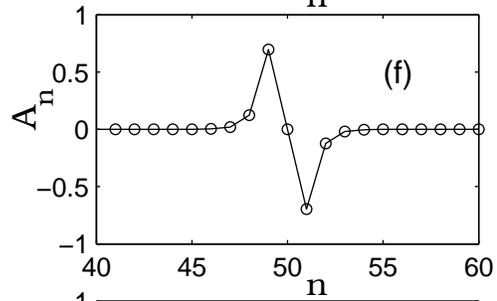
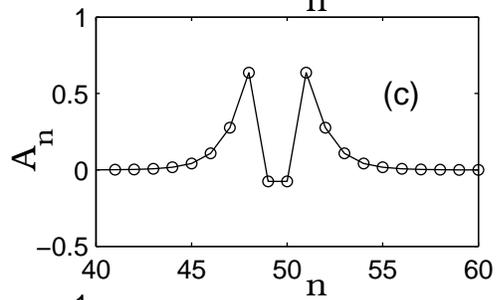
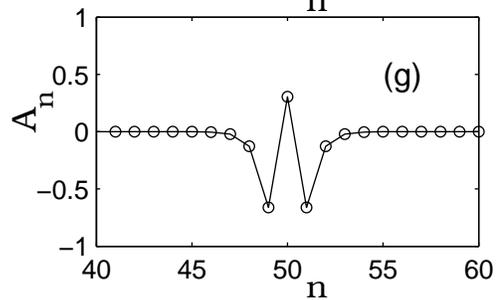
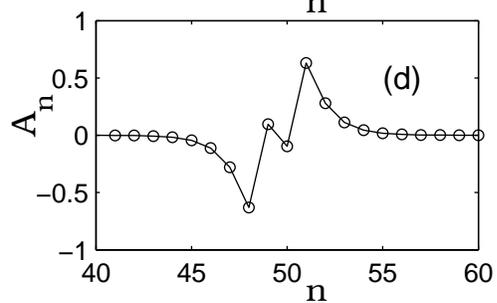
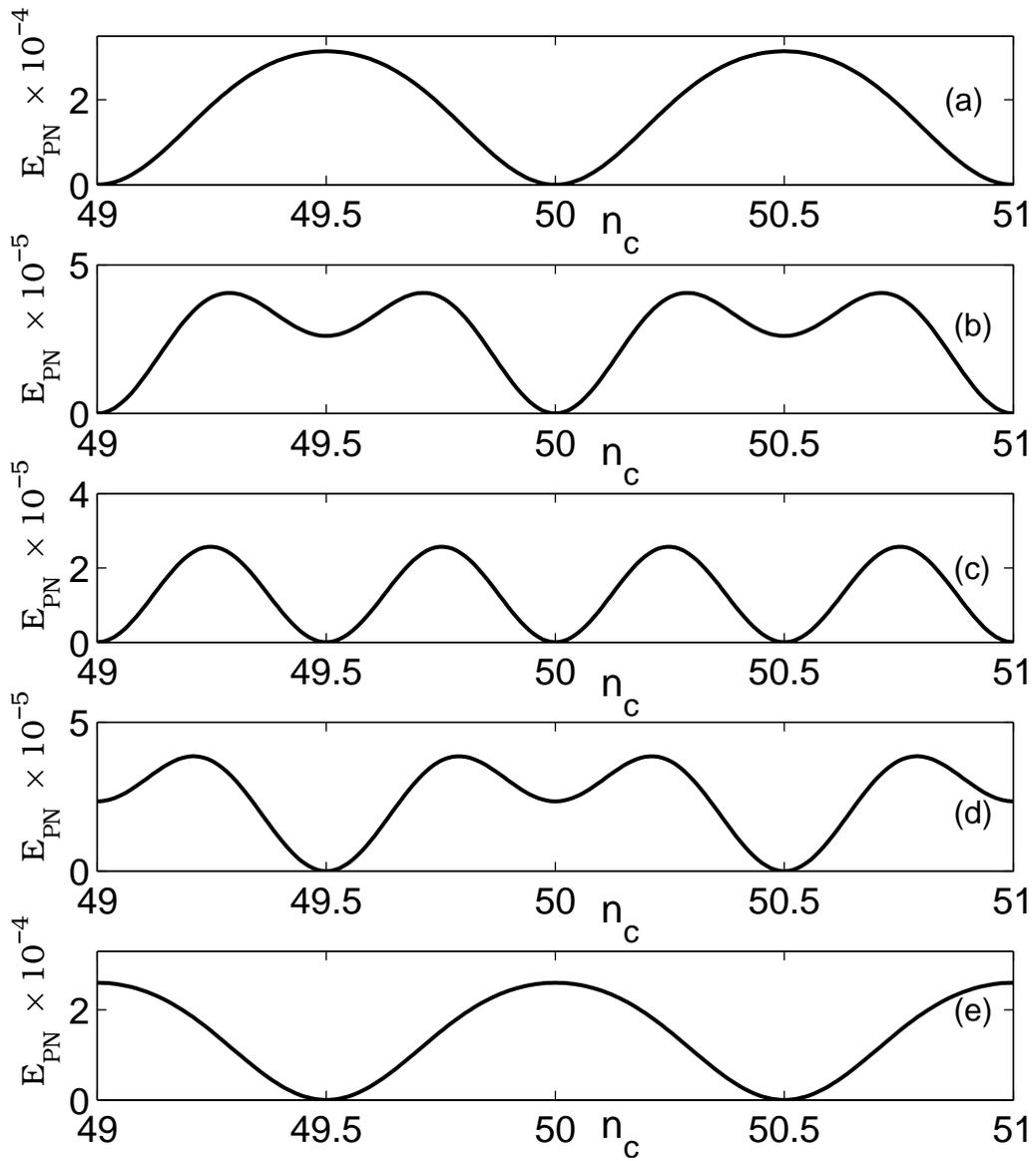
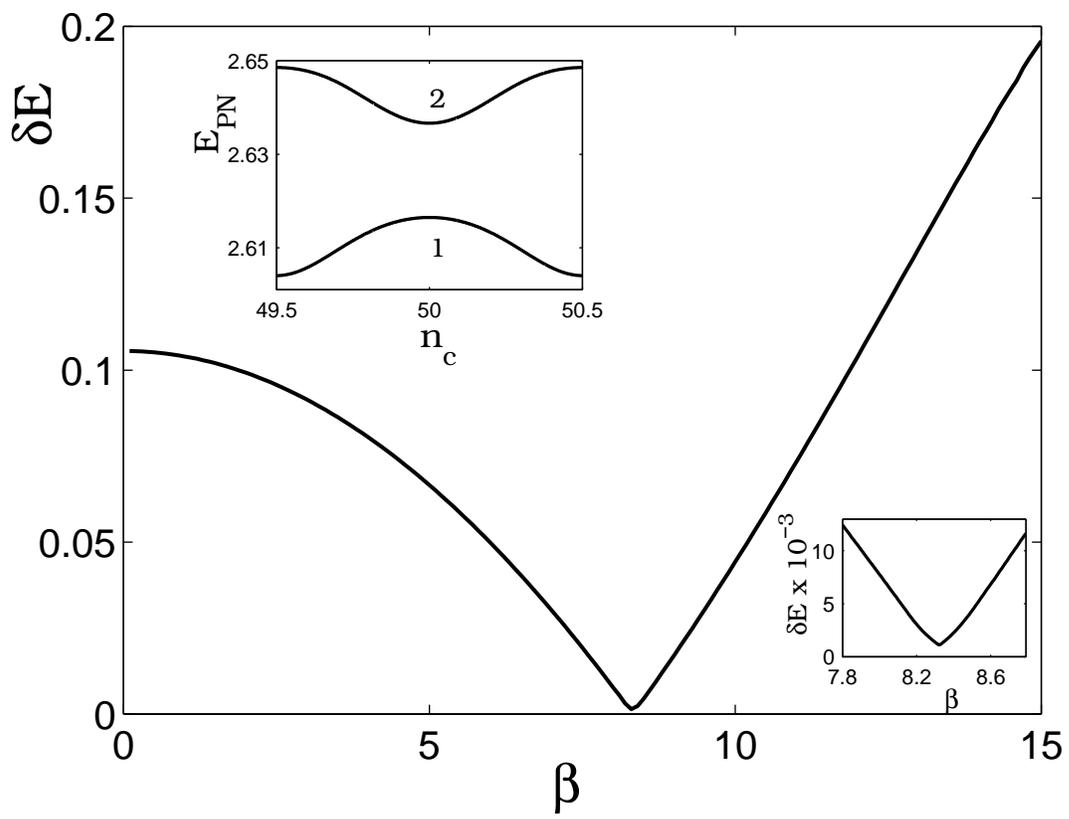
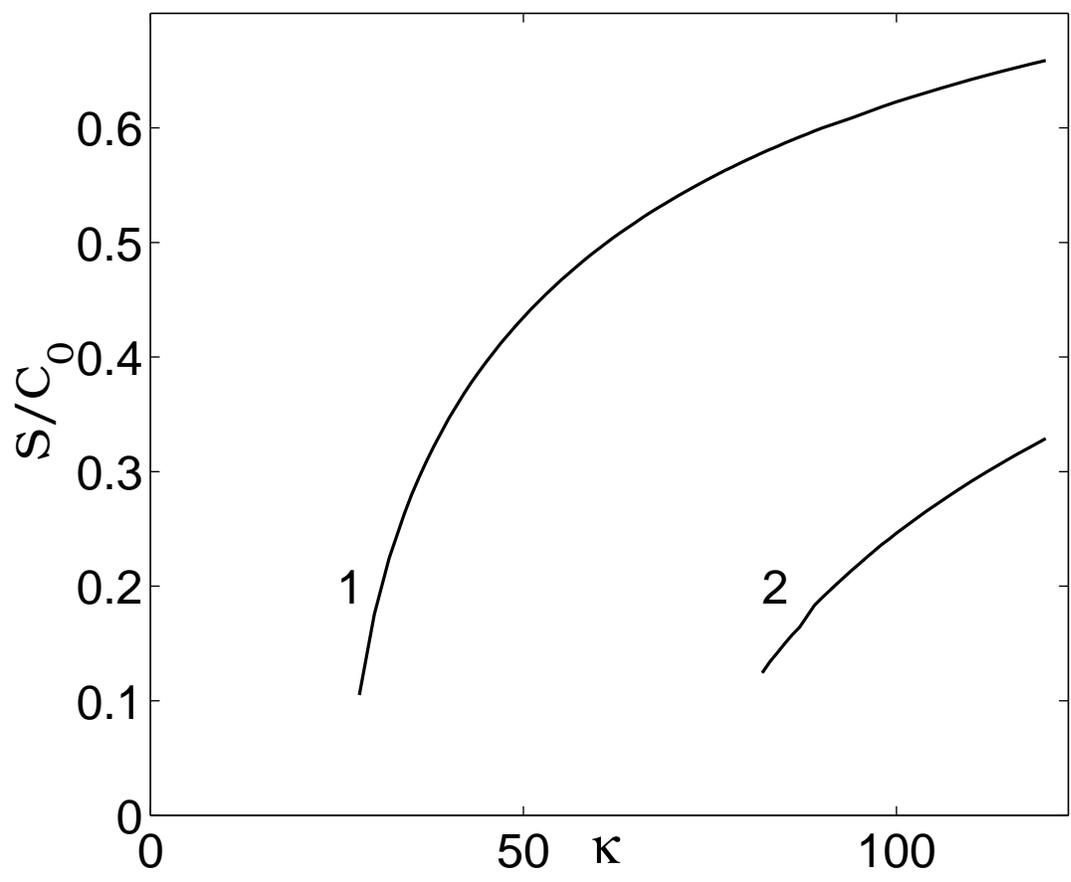
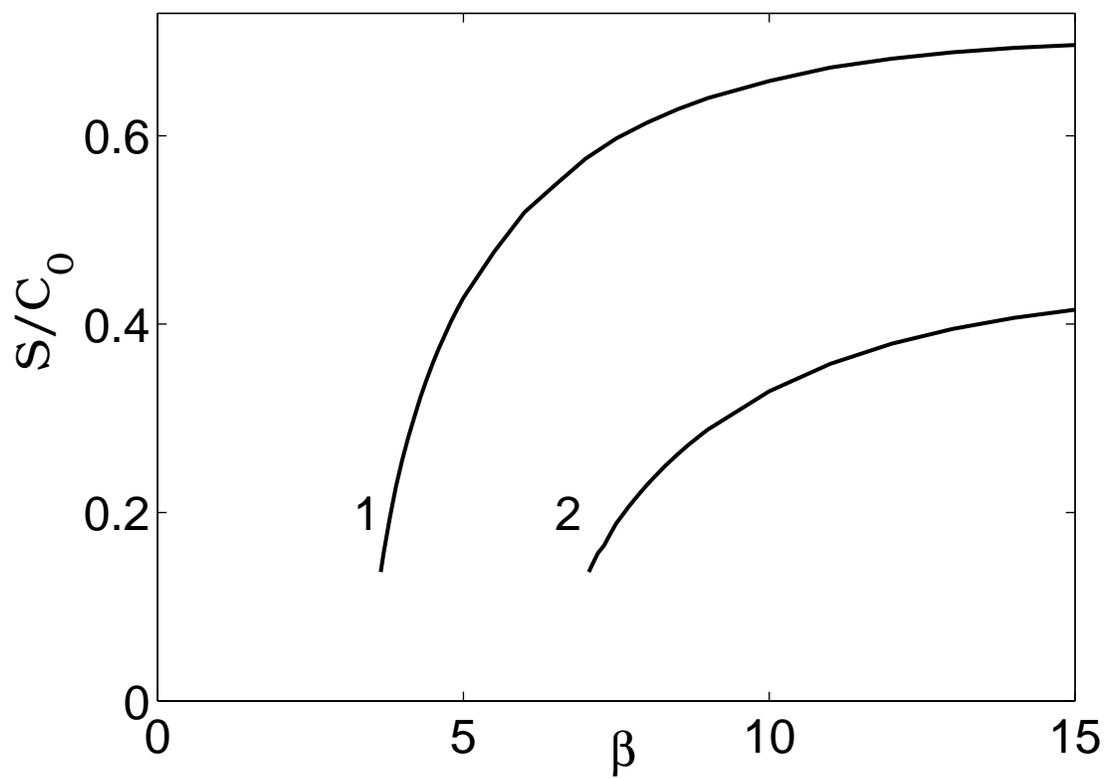
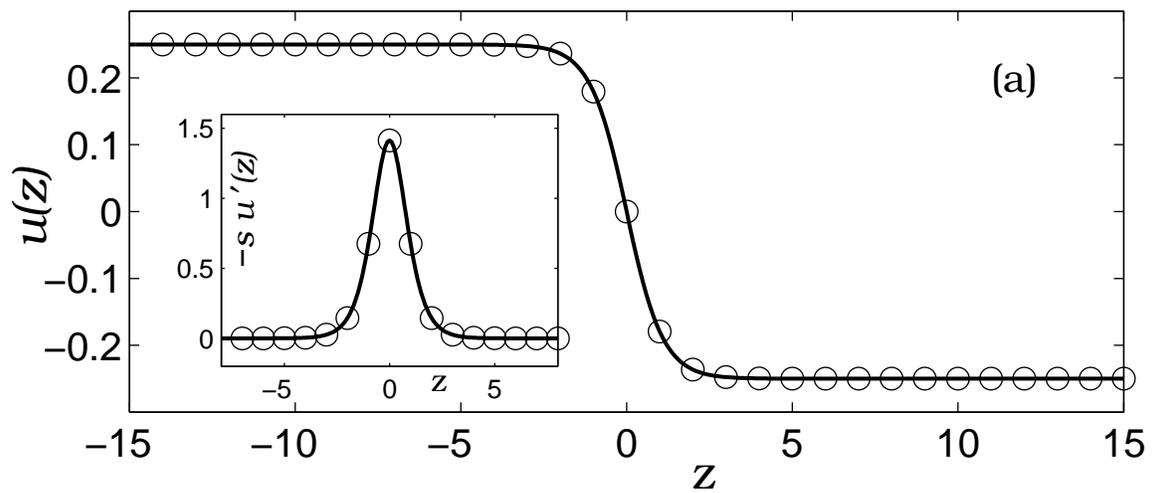
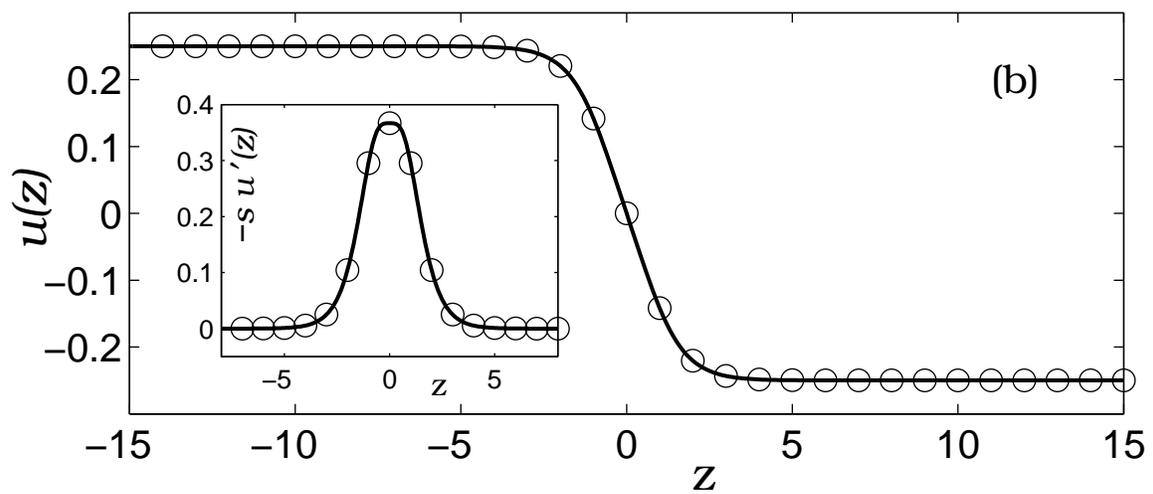
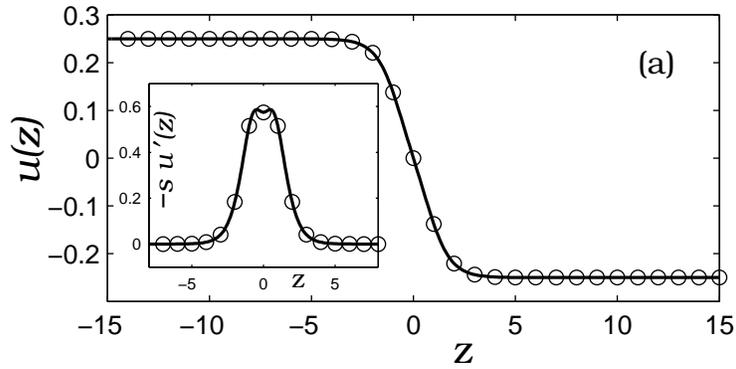
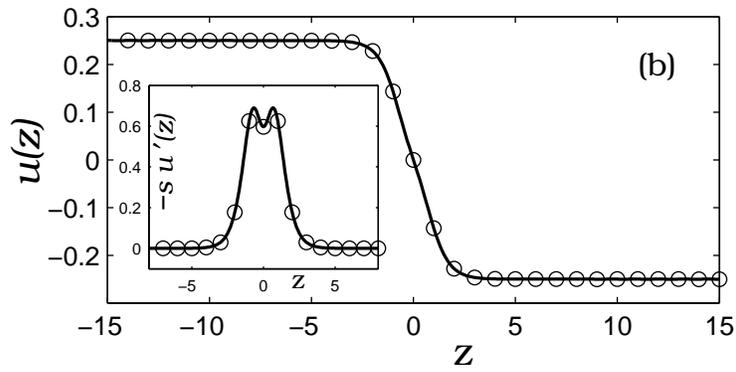
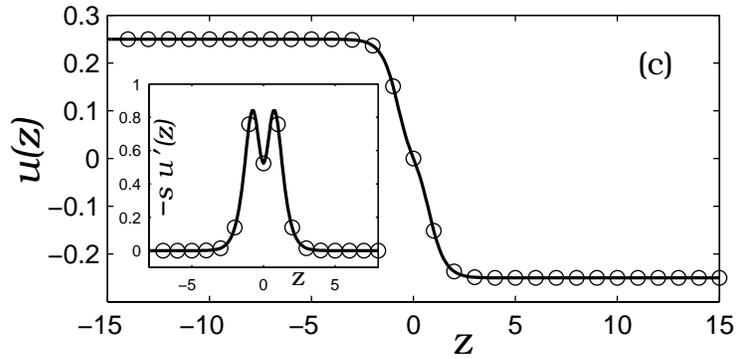
\end{document}